\documentclass[twoside,12pt]{article}
\usepackage{epsfig}

\def\Journal#1#2#3#4{{#1} {#2} (#4) #3 }

\def\NPA{{\em Nucl. Phys.} A}

\def\NPB{{\em Nucl. Phys.} B}

\def\PLB{{\em Phys. Lett.} B}

\def\PRL{\em Phys. Rev. Lett.}

\def\PRD{{\em Phys. Rev.} D}

\newcommand{\be}{\begin{equation}}
\newcommand{\ee}{\end{equation}}
\newcommand{\bea}{\begin{eqnarray}}
\newcommand{\eea}{\end{eqnarray}}

\begin{document}

\title{ \vspace{1cm} BBN with Late Electron-Sterile
Neutrino Oscillations \\ - The Finest Leptometer}
\author{D.\ Kirilova,$^{1}$\\
\\
$^1$Institute of Astronomy and NAO, BAS, Bulgaria\\
}
\maketitle
\begin{abstract}
A relic lepton asymmetry orders of magnitude bigger than the baryon
one may hide in the relic neutrino background. No direct theoretical
or experimental limitations on its magnitude and sign are known.
Indirect cosmological constraints exist ranging from $|L|<0.01$ to
$L<10$. We discuss  a BBN model with late electron-sterile neutrino
oscillations which is a fine leptometer - it is capable of feeling
extremely small relic lepton asymmetry - $|L|>10^{-8}$. This
sensitivity is achieved through the influence of such small $L$ on
the neutrino oscillations, suppressing or enhancing them,  and thus
changing the primordially produced $^4He$. The influence of $L$ on
nucleons freezing in pre-BBN epoch is numerically analyzed in the
full range of the oscillation parameters of the  model  and for $ L
\ge 10^{-10}$. The case of oscillations generated asymmetry by late
electron-sterile oscillations and its effect on primordial $^4He$ is
also briefly discussed.

\end{abstract}
\section{Introduction}

Lepton asymmetry of the Universe is usually defined as
$L=(N_l-N_{\bar{l}})/N_{\gamma}$, where $N_l$ is the number density
of leptons, and $N_{\bar{l}}$ of antileptons, while $N_{\gamma}$ is
the number density of photons. In case of equilibrium $\xi=\mu/T$,
the degeneracy parameter,
  is used instead as a qualitative measure  of the lepton asymmetry.

  It is traditionally
assumed that the lepton asymmetry is of the order of the baryon one,
which is measured precisely by different independent means (BBN, CMB
data) to be $\beta=(N_b-N_{\bar b})/ N_{\gamma} \sim 6.10^{-10}$.
However, actually
 big $L$ may
reside in the neutrino sector~\footnote{Universal charge neutrality
implies that the lepton asymmetry in the electron sector is $L_e\sim
\beta$.} and hence, $L$  may be many orders of magnitude larger than
$\beta$.  Therefore, $L$ is defined mainly by the sum of the
asymmetries in the different neutrino sectors $L\sim\sum L_{\nu_i}$.
Direct measurements of the lepton asymmetry magnitude and sign have
not been done yet.  There are available just indirect indications
and constraints.

Studying $L$ of the Universe is intriguing and important because of
many reasons. Just a short list of some of them reads:

\begin{itemize}

 \item A significant cosmological
effect of $L$ might be expected, having in mind that neutrinos are
abundant and large $L$ may be contained in the neutrino sector.

\item  Precise determination of
neutrino properties is of cosmological importance. As far as the
uncertainty of neutrino characteristics leads to large systematic
errors in the estimation of cosmological parameters obtained from
CMB data. Therefore, knowledge about $L$ and its nature would help
to determine them more precisely.

 \item Studying $L$ and its cosmological influence  provides,
 on the other hand, an opportunity to
 use cosmology as a probe of neutrino properties.

\item There exist different mechanisms of generation of $L$, and among them
   is the  natural possibility of amplifying the neutrino asymmetry by neutrino
active-sterile resonant oscillations in $CP$-asymmetric plasma of
the early Universe~\cite{FVPRD96,KC}.

\item Planck will soon be able to better constrain $L$, namely Planck
will measure the radiation content at CMB decoupling with a
precision $\delta N_{eff}=\pm 0.26$, i.e. it will be able to measure
with this precision the eventual $L$ contribution to the radiation
content. Besides, in case $\xi>2$  Plank will be able to detect
small neutrino mass of the order of $\sim 0.07$ eV because $L$
enhances the effect of the mass.

  \item Today it is reasonable to study $L$, also as a possible solution  of the
recently found cosmological preference or/and indication for
additional relativistic density~\cite{friend}. Namely,  recent
measurements of primordially produced $^4He$ $Y_p$ ~\cite{IT10} and
 CMB measurements of $Y_p$~\cite{WMAP7} point to an effective
number of the relativistic degrees of freedom at the BBN epoch
$N_{BBN}=3.8^{+0.8}_{-0.7}$, at the CMB formation epoch
$N_{CMB}=4.34^{+0.86}_{-0.88}$ at $68\%$, and on the basis of LSS
survey data $N_{LSS}=4.8^{+1.9}_{-1.8}$ at $95\%$, i.e. higher
values than previously estimated ones and higher than the
theoretically predicted standard value for 3 neutrino species
$N_{eff}=3.046$.

  \item Moreover, recent analysis of the combined neutrino oscillation data,
including LSND and MiniBoone requires 1 or 2 additional low mass
sterile neutrinos $\nu_s$, participating into oscillations with
flavor neutrinos with higher mass differences values, than the ones
required by solar and atmospheric neutrino oscillations experiments.
These light $\nu_s$ brought into equilibrium by oscillations with
active neutrinos may be a successful explanation of the excess
relativistic density. Besides, it is known that such active-sterile
oscillations in the resonant case may generate $L$ in the
CP-asymmetric plasma of the Universe.

  \item Knowledge about $L$ magnitude and sign is relevant  for cosmological
  models with leptonic domains, for inhomogeneous BBN models,
  baryogenesis through leptogenesis issues, etc.

\item Determining $L$ at BBN epoch would allow
testing  the assumption that sphalerons equilibrate lepton and
baryon asymmetries.

\item Particularly concerning neutrino oscillations: Due to the fact that
$L$ is capable of suppressing and inhibiting or enhancing neutrino
oscillations, determining $L$ will enlighten our knowledge about
their cosmological role.

\end{itemize}

In the next section lepton asymmetry effects on processes in the
Universe  and some of the based on them cosmological constraints on
$L$ are shortly discussed.  In section 3 the interplay between small
lepton asymmetry and electron-sterile neutrino oscillations $\nu_e
\leftrightarrow \nu_s$, effective after neutrino decoupling, is
described in terms of general kinetic equations for oscillating
neutrinos and  numerically analyzed. In section 4 the results of our
numerical analysis of the influence of relic $L$ and oscillations
generated $L$ on primordial production of $^4He$ in case of late
$\nu_e \leftrightarrow \nu_s$ oscillations are presented.

\section{Knowledge about Lepton Asymmetry}

Cosmic neutrino Background has not been detected yet, hence $L$ is
measured/constrained only indirectly through its effect on other
processes, which have left observable  traces in the Universe, like
the abundances of light elements produced in BBN,
Cosmic Microwave Background, LSS, etc.
(see for example the reviews ~\cite{Dolgov02} -- \cite{
Lesgourgues&Pastor}). Recently it was found possible to obtain an
information about $L$ from
  QCD transition, on the basis of $L$ effect at QCD epoch
~\cite{Schwarts&Stuke09}. Below we review the main cosmological
effects of $L$ and some of the recent cosmological constraints on
$L$ magnitude and sign.

\subsection{Lepton Asymmetry Effects}

{\bf A. Energy density increase.}

A well-known cosmological effect of $L$ is the increase of the
radiation energy density, which  is usually expressed in terms of
the increase of the effective number of the  relativistic degrees of
freedom
$$
\rho_r=\rho_{\gamma}+\rho_{\nu} = [1+7/8(4/11)^{4/3}
N_{eff}]\rho_{\gamma}
$$
where $\rho_{\gamma}$ and $\rho_{\nu}$ are the photon and neutrino
energy densities, correspondingly. In equilibrium $L$ may be
expressed as usual  through the chemical potential $\mu$ or
degeneracy parameter $\xi=\mu/T$:

$$L=1/12 \zeta(3) \sum_i T^3 _{\nu_i}/T^3 _{\gamma} (\xi^3 _{\nu_i} +
\pi^2 \xi_{\nu_i})$$

The increase of $N_{eff}$ due to $L$ is
$$
 \Delta N_{eff}=15/7[(\xi/\pi)^4+2(\xi/\pi)^2].
$$

 The increase of the
radiation density due to $L$ speeds up the Universe expansion
$H=(8/3\pi G \rho)^{1/2}$, delays matter/radiation equality epoch,
changes the decoupling temperature of neutrino, which on their turn
influence BBN, CMB and the evolution of the density perturbations,
i.e. formation of structures in the Universe.

Particularly well studied is $L$ effect on BBN. The increase of the
cooling rate of the Universe due to $L$, leads to earlier freezing
of the reactions governing neutron-to-proton ratio $n/p$ , i.e.
leads to higher freezing ratio $(n/p)_f$, which reflects in higher
$D$ and $^4He$ abundances.

{\bf B. Direct kinetic effect.}

Besides its dynamical effect, lepton asymmetry with a magnitude
$|L|>0.01$ in the $\nu_e$ sector exerts also a direct kinetic effect
on the n-p kinetics and on BBN, because the
 $\nu_e$  participates in the reactions interconverting
neutrons and protons.
In this case the effect on BBN and the outcome of the light elements
is $L$ sign dependent.

As is obvious, $L>0$ in the pre-BBN epoch would result into
reduction of $(n/p)_f$ and thus leads to light element
underproduction, while $L<0$ would lead to their overproduction.
Degenerate BBN has been thoroughly studied (see for example pioneer
papers ref.~\cite{degenerate}).

 An empirical
 formula, which provides a fairly good fit (see
 for example ref.~\cite{Simha&Steigman}), presents the dependence of the produced
primordially $^4$He, $Y_p$, on the discussed dynamical and kinetic
effect of $L$:
$$
Y_p\sim (0.2482\pm 0.0006)+0.0016\eta_{10}+0.013\Delta
N_{eff}-0.3\xi_{\nu_e}
$$


{\bf C. Indirect kinetic effect due to asymmetry-oscillations
interplay}

Small $L$, $|L|<< 0.01$, that has negligible {\it A} and {\it B}
effects, in case of late $\nu_e \leftrightarrow \nu_s$ oscillations
may considerably influence oscillating $\nu_e$, namely change its
evolution, number density, energy distribution, oscillation pattern
and thus through $\nu_e$ influence BBN kinetics~\cite{KC,NPB98}. The
effect of small relic $L$ and  nonresonant $\nu_e \leftrightarrow
\nu_s$ oscillations effective after neutrino decoupling on BBN has
been first studied in ref.~\cite{NPB98}. It was found that $L<
10^{-7}$ is destroyed by oscillations, while $L \ge 10^{-7}$ {\it
may enhance or suppress oscillations}
and through them influence primordially produced elements (see also
ref. ~\cite{Verbier}.~\footnote{ The possibility of $L$ to suppress
neutrino oscillations was discussed in the case of oscillations with
larger mass differences first in ref.~\cite{FV}.}

On the other hand active-sterile oscillations may  {\it induce
neutrino-antineutrino asymmetry growth}
  during the resonant transfer of
neutrinos~\cite{FVPRD96,KC,shi,NPB00}. The case of weak mixing and
relatively big mass differences $\delta m^2> 10^{-4}$ eV$^2$ was
discussed first in  ref.~\cite{FVPRD96}, while the case of asymmetry
generated at relatively big mixing and small mass differences
$\delta m^2< 10^{-7}$ eV$^2$  was first found in ref.~\cite{KC}.
This dynamically produced asymmetry exerts back effect on
oscillating neutrino  and changes its oscillation pattern.
~\footnote{The case of fast oscillations generating high enough $L$
that exerts direct kinetic effect {\it B} on BBN was discussed in
many publications (see for example ref.~\cite{FV97}).} There exists
another possibility: when $L$ growth is not high enough to have a
direct $L$ kinetic effect on the synthesis of light elements, it can
effect indirectly BBN through its effect on oscillating neutrinos.
In the case of resonant
 $\nu_e \leftrightarrow \nu_s$ oscillation effective after neutrino
 decoupling
 neutrino-antineutrino asymmetry is amplified by not
more than 5 orders of magnitude from an initial value of the order
of the baryon one~\cite{NPB00}. Oscillations generated asymmetry
{\it suppresses oscillations at small mixing angles}, leading to
noticeable decrease of $^4He$ production at these mixing angles.
The effect of small $L$  generated by oscillations on $^4He$
abundance and on cosmological constraints on oscillations was
analyzed in ref.~\cite{KC,NPB00,Verbier}.

In conclusion, very small asymmetries $10^{-7}<L\ll 0.01$ , either
relic or produced
 in active-sterile oscillations,  may considerably influence oscillating electron neutrino
  and through it $Y_p$ and BBN.
In the next sections we study numerically this special case of
influence of small $L$ on oscillating neutrino and on BBN and show
that BBN produced $^4He$ feels extremely small $L$ and represents
now the finest known "leptometer". For comparison, in the next
subsection we present first the available cosmological constraints
on $L$ coming from BBN with flavor oscillations present.

\subsection{Lepton Asymmetry Constraints}

 {\it At present BBN provides the most stringent constraints on L.}
 There exist numerous papers on the subject.
I list here only the results of more recent work. For more
information and reference of earlier papers see the review paper
~\cite{Dolgov02}.


In case of equilibration of the neutrino oscillation degeneracies
due to flavor oscillations before BBN the limit on $L$ in the muon
and tau neutrino sector strengthens. Then BBN constraint reads
~\cite{dolgov&petcov}
 $|\xi_{\nu}|<0.1$.
A more recent study for that case gives the constraint :
$-0.04<\xi_{\nu_e}<0.07$ ~\cite{Serpico&Raffelt}. For such a small
$L$ the expantion rate remains practically standard because $\delta
N_{eff} ~ 10^{-3}$.

However,  it was realized later that the equilibration of the
chemical potentials before BBN depends on the value of the yet
unknown mixing $\theta_{13}$. ~\cite{SerpicoPinto&Raffelt}. Besides,
since relic $L$ is capable to suppress or enhance oscillations,
depending on its value and the values of oscillation parameters, $L$
itself may play the role of an inhibitor or a catalyzer of the
equilibration. Hence, different possibilities for the chemical
potential in different neutrino flavors still may have place.

The analysis on the basis of BBN and $D$ and $^4He$ abundance and
CMB/LSS constraints on baryon-to photon value, provided restrictive
constraints on the neutrino degeneracy~\cite{Simha&Steigman}. Namely
the following constraints were derived for
$N_{eff}=3.3^{+0.7}_{-0.6}$ and different possibilities for the
chemical potentials: in case  $\xi_{\nu_e}\ne
\xi_{\nu_{\mu}}=\xi_{\nu_{\tau}}$   $\xi_{\nu}<2.3$ corresponding to
$L<5$;  in case $\xi_{\nu_e}=\xi_{\nu_{\mu}} \ne \xi_{\nu_{\tau}}$
$\xi_{\nu_{\tau}}<4$  $L<7.6$, while in case
$\xi_{\nu_e}=\xi_{\nu_{\mu}}=\xi_{\nu_{\tau}}$ $0.01<\xi_{\nu}<0.1$
and $L<0.07.$ In the last case practically the rate of expansion
does not change, and the small dynamical effect of $L$ corresponding
to $\Delta N_{eff}\sim 0.03$ is undetectable by BBN and CMB
~\cite{PastorPinto&Raffelt}.

 {\it CMB and LSS provide much looser bounds:} $L$ modifies the power spectra of
radiation and matter
~\cite{Lesgourgues&Pastor},\cite{Popa&Vasile08}. However,  today's
sensitivity of CMB and LSS data  does not allow to probe different
flavors. They feel only  the change in the total density, i.e.
$\delta N_{eff}$.
 CMB data (including
WMAP 5 years data results) combined with LSS data puts the bound:
$\xi_{\nu}<0.7$ and $L<0.6$ at $2\sigma$ level.

The WMAP5 data combined with the data on BBN produced helium-4
provides more stringent bounds, namely: $-0.04<\xi_{\nu}<0.02$ in
case of equilibration, while otherwise  $-0.03<\xi_{\nu_e}<0.13$,
$|\xi_{\nu_{\mu,\tau}}|<1.67$ ~\cite{Shiraishi09}. See however the
recent work
~\cite{Christel} based on the new data on $Y_p$ and WMAP7, which
allows considerable relaxation of the constraints, namely
$-0.14<\xi_{\nu_e}<0.12$.


In conclusion, depending on the different combinations of
observational data sets used and the assumed uncertainties,
cosmology provides an upper bound for $L$ in the range
$|L_{\nu_{\mu,\tau}}|<10^{-2}-10$  and $|L|<0.01-0.2$. These values
are many orders of magnitude larger than the baryon asymmetry value.

In the next sections we analyze the effect of small $L$ on BBN with
late neutrino oscillations and present a possibility for much
sensitive leptometer capable to constrain $L$ values closer to the
baryon one.  Namely, we describe a model of BBN with late
electron-sterile neutrino oscillations, which can 'measure' $L$ as
small as $L=10^{-8}$.

\section{Small Lepton Asymmetry and Late  Neutrino Oscillations}

Here we present the results on the interplay between small
asymmetries $L<<0.01$ and late $\nu_e \leftrightarrow \nu_s$
oscillations in the early Universe and their effect on BBN. This
study continues and broadens the analysis of refs.~\cite{NPB98} and
~\cite{NPB00}.

The effect of large $L$ either previously existing or produced by
oscillations, which effect directly the kinetic of BBN has been
studied in numerous papers.
 The effect of small relic asymmetry was not so thoroughly studied.
 Its effect on primordial
$^4$He abundance was first analyzed for hundreds of $\delta
m^2-\theta$ combinations in refs.~\cite{NPB98} and ~\cite{KC,NPB00}.

We  have  analyzed electron-sterile neutrino oscillations
$$
\nu_1=\cos(\theta) \nu_e + \sin(\theta) \nu_s
$$
$$
\nu_2=-\sin(\theta)\nu_e + \cos (\theta)\nu_s,
$$
where $\theta$ is the mixing angle, $\nu_1$ and $\nu_2$ are the
Majorana particles with masses $m_1$ and $m_2$. The sterile neutrino
$\nu_s$ is not having the usual weak interactions, and is assumed to
have decoupled much earlier than the flavor neutrinos, hence its
density is much lower than the density of electron neutrino
$n_{\nu_s}<<n_{\nu_e}$. The case of neutrino oscillations effective
after active neutrino decoupling $\delta m^2 \sin^4 2\theta \le
10^{-7}$ eV$^2$ is studied in detail.

For that specific case of  small $L$ both the dynamical effect {\it
A}, discussed in the previous section, and the direct kinetic effect
{\it B} on nucleon kinetics are negligible. We will discuss the
asymmetry effect {\it C} on oscillating neutrinos and through them
on BBN.

In the case of late oscillating active-sterile neutrinos with relic
or generated in oscillations $L$ the energy distribution of
neutrinos may be strongly distorted from the equilibrium Fermi-Dirac
form~\cite{NPB00}. Hence, a precise account for the energy spectrum
distortion of the degenerate oscillating neutrinos is necessary to
reveal the effect of small lepton asymmetry. Particularly the
capability of small relic $L$ to enhance oscillations has
essentially spectral character~\cite{NPB98} and requires a precise
kinetic approach, provided in the numerical analysis, described
below.

We have studied two different cases of $L$, namely initially present
at the neutrino decoupling epoch, called further on relic $L$, and
dynamically generated $L$ by oscillations.

\subsection{Evolution of Oscillating Neutrino in Presence of Lepton Asymmetry}
We have used a  self consistent
 numerical analysis of
the kinetics of the oscillating
 neutrinos,
the  nucleons freeze-out and the asymmetry evolution  for the
analysis of lepton asymmetry role in BBN. Kinetic equations for
neutrino density matrix and neutron number densities in momentum
space are used to describe the evolution of the system of
oscillating neutrinos in the high temperature Universe, following
the approach of ref.~\cite{KC}.

\begin{eqnarray*}
\partial \rho(t) / \partial t &=&
H p_\nu~ \left(\partial \rho(t) / \partial p_\nu\right) +\\
&&+ i \left[ {\cal H}_o, \rho(t) \right] +i \sqrt{2} G_F \left({\cal
L} - Q/M_W^2 \right)N_\gamma \left[ \alpha, \rho(t) \right]
+ {\rm O}\left(G_F^2 \right),\\
\partial\bar\rho(t) / \partial t&=&
H p_\nu~ \left(\partial \bar\rho(t) / \partial p_\nu\right) +\\
&&+ i \left[ {\cal H}_o,\bar\rho(t) \right] +i \sqrt{2} G_F
\left(-{\cal L} - Q/M_W^2 \right)N_\gamma \left[ \alpha, \bar\rho(t)
\right]
+ {\rm O}\left(G_F^2 \right),\\
\ \\
\partial n_n / \partial t
&=& H p_n~ \left(\partial n_n / \partial p_n \right) +\\
&& + \int {\rm d}\Omega(e^-,p,\nu) |{\cal A}(e^- p\to\nu n)|^2
\left[n_{e^-} n_p (1- {\rho_{LL}}) - n_n {\rho_{LL}}
(1-n_{e^-})\right]\\
&& - \int {\rm d}\Omega(e^+,p,\tilde{\nu}) |{\cal A}(e^+n\to
p\tilde{\nu})|^2 \left[n_{e^+} n_n (1- {\bar{\rho}_{LL}}) - n_p
{\bar{\rho}_{LL}} (1-n_{e^+})\right].
\end{eqnarray*}
\ \\

\noindent $\alpha_{ij}=U^*_{ie} U_{je}$,
 $\nu_i=U_{il}\nu_l (l=e,s)$. ${\cal H}_o$ is the free neutrino Hamiltonian.
 $Q$ arises as an $W/Z$ propagator effect,$Q \sim E_\nu~T$.
 {${\cal L} \sim 2L_{\nu_e}+L_{\nu_\mu}+L_{\nu_\tau}$,
$L_{\mu,\tau} \sim (N_{\mu,\tau}-N_{\bar{\mu},\bar{\tau}})/
N_\gamma$
 $L_{\nu_e} \sim \int {\rm d}^3p(\rho_{LL}-\bar{\rho}_{LL})/N_\gamma$.

The first two equations describe the evolution of neutrino and
antineutrino ensembles. They provide a {\it simultaneous account} of
the different competing processes: expansion (first term), neutrino
oscillations (second term), neutrino forward scattering and weak
interaction processes.  The number densities of nucleons and
electron neutrino were assumed the equilibrium ones. The sterile
state was assumed empty at the time of decoupling of the electron
neutrino.~\footnote{ the case of non-zero population of the sterile
neutrino state was considered in detail for $L=10^{-10}$ case in
refs.~\cite{D04,D07,DP06}}

Due to the non-zero $L$ term the equations are coupled
integro-differential  and the numerical task is much complicated
than in the case of zero $L$. Besides, $L$ term leads to different
evolution of neutrino and antineutrino due to the different sign
with which it enters their equations. (The case of $L~\beta$
corresponds to negligible $\cal L$ term  in the potential when the
evolution of the neutrino and antineutrino density matrices is
identical.

We studied  numerically the evolution of  neutrino ensembles,
evolution of  $L$,  and also the evolution of nucleons and $L$ role
during pre-BBN epoch for a broad range of oscillation parameters and
$10^{-10}<L<0.01$ for the case of a relic $L$. For the case of
oscillations generated $L$ its initial value was assumed $L~\beta$.
The numerical analysis was provided for the temperature range [$0.3$
MeV, $2$ MeV] and the full set of oscillation parameters of the
electron-sterile oscillations model, and with higher accuracy than
in previous studies. We have calculated precisely neutron to
nucleons freezing ratio $X_n^f=n_n^f/(n_n+n_p)^f=f(\delta m^2,
\sin^22\theta, L)$ which is essentially influenced by oscillations
and $L$. The primordially produced $^4He$ was estimated from it.

Active-sterile oscillations proceeding after neutrino decoupling
produce $\nu_s$ at the expense of active neutrino and thus $\delta
N_eff$ does not change. However in that case late oscillations for a
wide range of values of oscillations parameters and $L$, strongly
distort neutrino energy spectrum. Therefore, a precise description
of neutrino momenta distribution is necessary.
In this analysis we have used between 1000 and 5000 bins  to
describe neutrino spectrum distribution in the non-resonant neutrino
oscillations case, and up to 10 000  in the resonant case.
       Depending on the oscillation parameters and
$L$ values, the following interplay between $L$ and oscillations can
be observed: relatively large $L$ suppress oscillations, smaller $L$
lead to their resonant enhancement.  On the other hand, resonant
oscillations also are capable to amplify $L$.  In the following
section we present the results of  a detail numerical study of this
interplay on BBN produced $^4He$.

\section{Lepton Asymetry, Neutrino Oscillations and BBN. The Results}

To study the lepton asymmetry effect on BBN we have provided a
detail numerical analysis of the influence of $L$ on $Y_p$, because
primordially produced $^4He$ is highly sensitive to the nucleons
kinetics during the pre-BBN epoch and besides, it is the most
precisely measured element among light elements synthesized during
BBN. A recent measurement of $Y_p$ was provided on the basis of 93
spectra of 86 low redshift HII regions~\cite{IT10}.

\subsection{Oscillations Generated Lepton Asymmetry and BBN}

In the analyzed oscillations case the evolution of $L$ is dominated
 by neutrino oscillations and typically $L$  has rapid oscillatory behavior:
it oscillates and changes sign.
We have determined numerically the region of parameter space for
which noticeable generation  of LA is possible. A good approximation
is $|\delta m^2|\sin^4 2\theta \le 10^{-9.5}$ eV$^2$. The maximal
possible growth of $L$ is by 4 -5 orders of magnitude. The
instability region and the magnitude of $L$ are close to the  bounds
existing in literature for other oscillation models~\cite{dolgL}.

 Precise description of the distribution of the neutrino momenta
 was found  extremely important
   for the correct determination of $L$ evolution  in the resonant oscillations case.
In some cases increasing  the resolution of momentum space  leads to
changes of the oscillatory character of $L$  and diminishes $L$
amplitude. `\footnote{As a rule in these cases  the evolution of the
neutrino ensembles is strongly distorted from the expected
behavior.} This observation is in accordance with the  studies of
other authors in other parameter regions corresponding to smaller
$\theta$ and bigger $\delta m^2$~\cite{Bari&Foot00}. Further
analysis is required to decide if the oscillatory behavior and
strong asymmetry growth is induced by numerical error.  This
observation revives the puzzle: Is the asymmetry growth due to lack
of numerical accuracy?!


$L$ changes energy spectrum distribution and the number densities of
electron neutrinos  from standard BBN case. This influences the
kinetics of nucleons
   during BBN and changes the production of light elements.

   We have precisely followed the evolution of nucleons in the
   presence of electron-sterile neutrino oscillations in
   the pre-BBN period for different sets of
   oscillation parameters
and different values of $L$. The production of  $^4He$
   was numerically calculated and compared to the BBN value without
   asymmetry growth  account.


\begin{figure}[htb]
\begin{center}
\begin{minipage}[t]{14 cm}
\epsfig{file=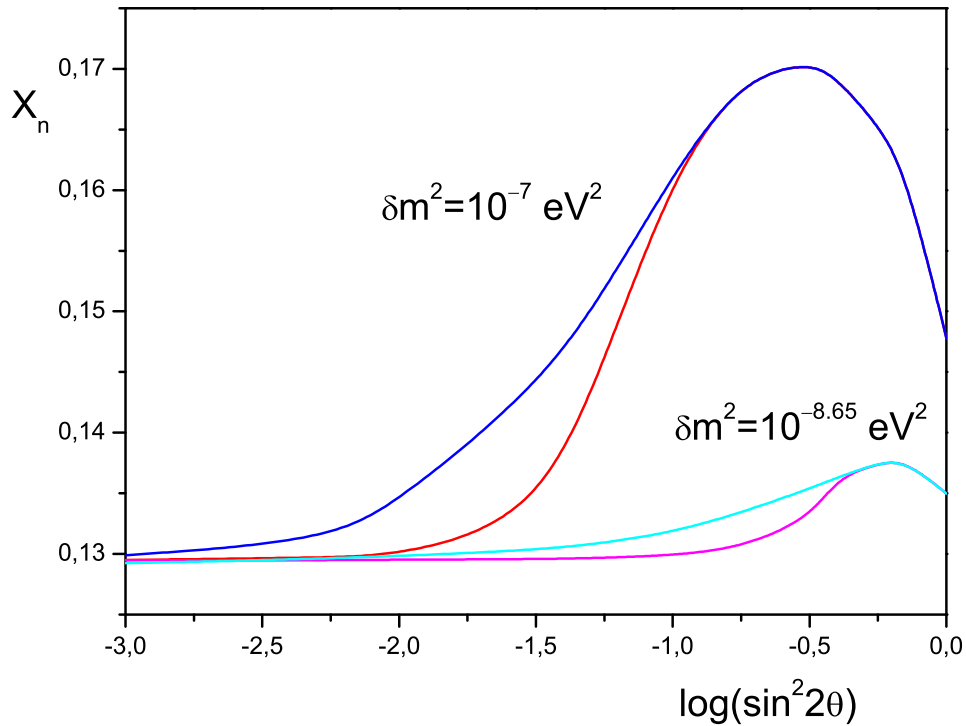,scale=0.7}
\end{minipage}
\begin{minipage}[t]{16.5 cm}
\caption{The dependence of the frozen neutron number density
relative to nucleons  on the mixing in case of the account of
asymmetry growth (red curves) and in case without asymmetry growth
account for two different mass differences $\delta m^2=10^{-8}$
eV$^2$ and $\delta m^2=10^{-7}$ eV$^2$.
 \label{fig1}}
\end{minipage}
\end{center}
\end{figure}

Fig.~\ref{fig1}  illustrates the typical behavior of the frozen
neutron number density relative to nucleons when increasing the
mixing in case of asymmetry growth (red curves) and in case without
asymmetry growth account for two different mass differences. The
asymmetry growth takes place at smaller mixing angles when
increasing $\delta m^2$. Then {\it due to the asymmetry growth the
production of} $X_n$ (correspondingly $Y_p$) {\it decreases at small
mixing}.  The effect of the asymmetry growth 
on helium production is always towards decreasing of the
 caused by oscillations overproduction of $^4He$, leading to a
relaxation of BBN constraints at small mixings.



In case of the small lepton asymmetry values, discussed here,   BBN
constraints on neutrino oscillations may noteably change
~\cite{NPB00,D07,DP06}. They  become less stringent at small mixing
angles (where the growth of asymmetry takes place).

Thus, the analysis has proven that  BBN  is sensitive to the
oscillations generated asymmetry, which usually grow not more than 5
orders of magnitude and are small
 $|L|<10^{-5}$.

\subsection{Initial asymmetry, oscillations and BBN}

Numerical analysis of  $Y_p(\delta m^2, \theta, L)$ dependence has
been provided for the entire range of mixing parameters of the model
and relic $L \ge 10^{-10}$.
 Small $L$,  $10^{-8}<L<<0.01$, that do not  effect directly BBN kinetics,
influence indirectly  BBN via oscillations in agreement with
previous analysis~\cite{NPB98}.

    The calculated $^4He$ production  dependence on oscillation parameters
    and on $L$  shows that, in case of  neutrino oscillations: i)
 BBN  can feel extremely small $L$: down to $10^{-8}$. ii) Large enough
 $L$
 change primordial production
of $^4He$ by enhancing or suppressing oscillations. Depending on
oscillation values
    $L \ge 10^{-7}$ may enhance oscillations, while $L>0.1 (\delta
     m^2/{\rm eV}^2)^{2/3}$ may
    suppress oscillations, and  asymmetries as big as
    $L>(\delta m^2/{\rm eV}^2)^{2/3}$
    inhibit oscillations.
 $L$ enhancing
 oscillations leads to a higher production of $Y_p$. $L$  suppressing oscillations
 decreases $Y_p$ overproduction by oscillations. $L$ bigger
than $10^{-4}$ leads to a total suppression of oscillations, i.e. to
the standard BBN yield of $Y_p$, without oscillations.

\begin{figure}[htb]
\begin{center}
\begin{minipage}[t]{8 cm}
\epsfig{file=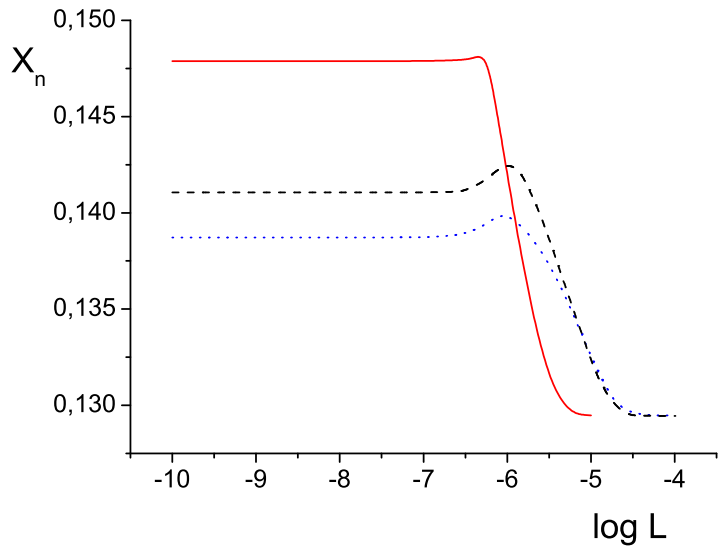,scale=0.5}
\end{minipage}
\begin{minipage}[t]{16.5 cm}
\caption{Frozen neutron number density relative to nucleons  as a
function of the relic initial lepton asymmetry for $\delta
m^2=10^{-7}$ eV$^2$. The solid curve corresponds to maximal mixing,
the dashed curve to $\sin^22\theta=10^{-0.05}$ and the dotted curve
to $sin^22\theta=10^{-0.1}$. \label{fig2}}
\end{minipage}
\end{center}
\end{figure}

In Fig.2 the dependences of $X^f_n$ on relic $L$ for different
mixings (to the left)  and different  mass differences are
presented. For $L$ smaller than $\sim 10^{-7}$  $X^f_n$ keeps
unchanged  from the case without $L$. The higher the mixing - the
higher is the overproduction of He-4 due to oscillations. Increasing
further $L$ for fixed oscillation parameters leads first to an
increase of helium production, corresponding to the region of
parameters space where $L$ enhances oscillations, and then to a
decrease of helium production, corresponding to big $L$ suppressing
oscillations, and hence to less $Y_p$ overproduction caused by
oscillations. At some critical $L$ value defined by the concrete set
of oscillation parameters $L_c(\delta m^2, \theta)$ the produced
helium reaches its standard BBN value - i.e. $L$ has stopped the
oscillations. As is illustrated in the figure, the width of the
enhancement region and the height of the overproduction peak is
sensitive to the mixing. Bigger values (up to about an order of
magnitude)for $L_c$ are necessary to inhibit oscillations when
decreasing the mixing.
\begin{figure}[htb]
\begin{center}
\begin{minipage}[t]{8 cm}
\epsfig{file=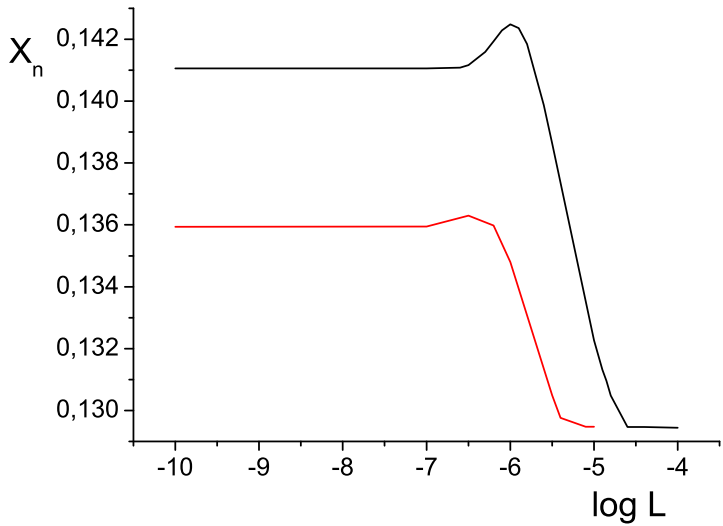,scale=0.5}
\end{minipage}
\begin{minipage}[t]{16.5 cm}
\caption{Frozen neutron number density relative to nucleons
dependence on the initial asymmetry for $\sin^22\theta=10^{-0.05}$
and two different mass differences $\delta m^2=10^{-8}$ eV$^2$
(lower curve) and $\delta m^2=10^{-7}$ eV$^2$ (upper curve).
\label{fig3}}
\end{minipage}
\end{center}
\end{figure}

Fig.3 illustrates the dependence of $X_n$ on the initial asymmetry
value for a fixed mixing, namely $\sin^22\theta=10^{-0.05}$ and
different mass differences $\delta m^2 = 10^{-8}$ eV$^2$ and $\delta
m^2= 10^{-7}$ eV$^2$. The enhancement peak due to $L$ is more
clearly expressed for higher mass differences, and $L_c$ is bigger
for bigger mass differences.
$L$ bigger than $\sim 10^{-4}$ leads to a total suppression of
oscillations effect on BBN for late oscillations studied here  and
hence, eliminates the BBN bounds on oscillation parameters. In that
case instead the following approximate bound holds: $\delta m^2/{\rm
eV}^2<L^{3/2}$ .

Depending on its value, relic $L$  may also change  BBN  bounds: It
 relaxes them at large  mixings and strengthens  them at small
mixings, as illustrated in ref.~\cite{NPB98}

\begin{figure}[htb]
\begin{center}
\begin{minipage}[t]{8 cm}
\epsfig{file=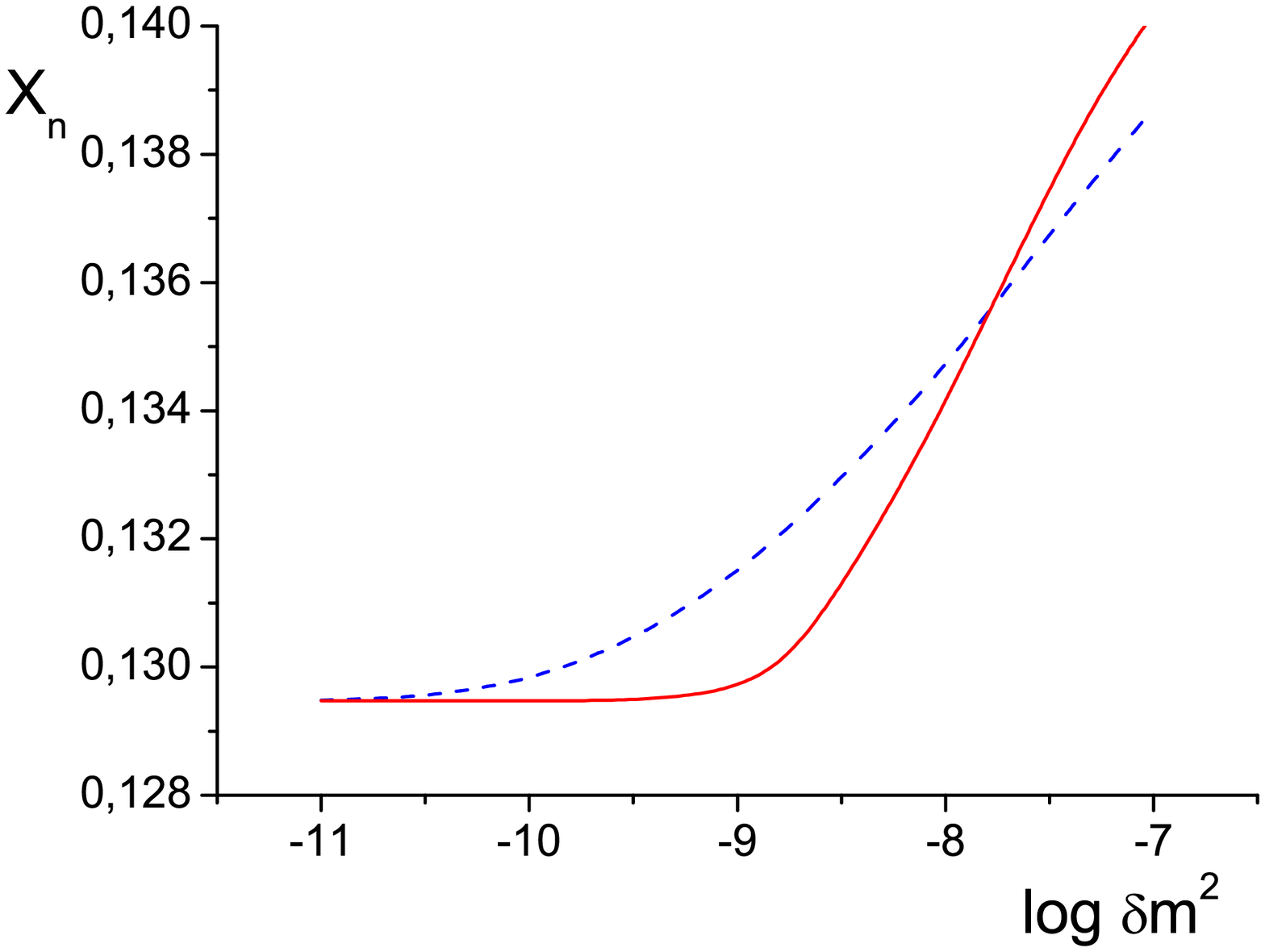,scale=0.5}
\end{minipage}
\begin{minipage}[t]{16.5 cm}
\caption{The dependence of the frozen neutron number density
relative to nucleons on the mass differences at
$\sin^22\theta=10^{-0.1}$ and for two different initial lepton
asymmetries $L=10^{-10}$ (the dashed curve) and $L=10^{-6}$ (solid
curve). \label{fig4}}
\end{minipage}
\end{center}
\end{figure}

The next Fig.4 presents the dependence of $X_n$ on the mass
differences at a fixed non maximal mixing angle and for two
different initial $L$. As illustrated,  higher initial $L$ leads to
an increase of helium production at bigger mass differences, and
reduces helium production at smaller mass differences.
Correspondingly, increasing $L$ at a fixed mixing leads to
relaxation of the bounds at small mass differences and strengthens
them fat big mass differences.

\begin{figure}[htb]
\begin{center}
\begin{minipage}[t]{8 cm}
\epsfig{file=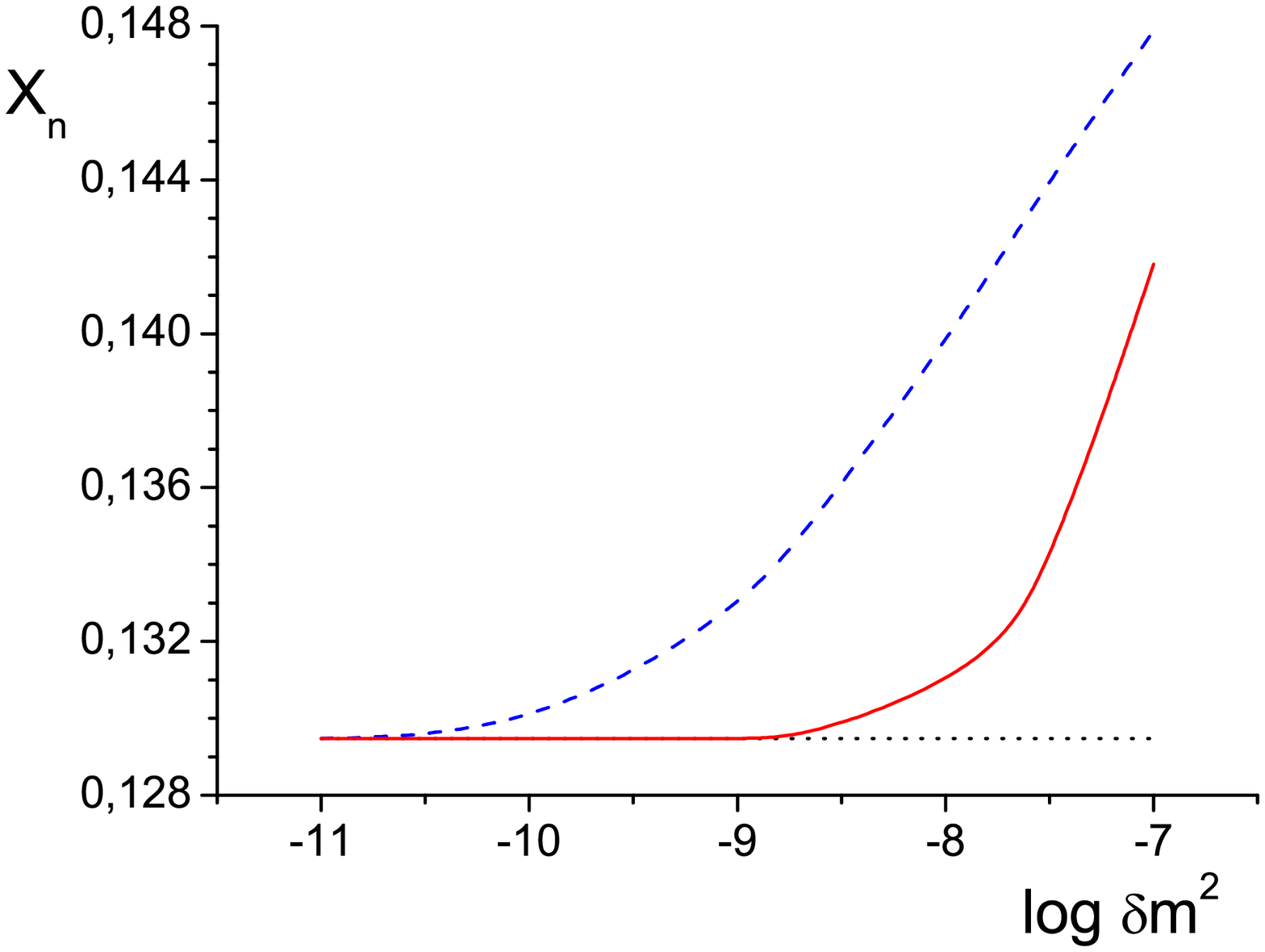,scale=0.5}
\end{minipage}
\begin{minipage}[t]{16.5 cm}
\caption{The dependence of the frozen neutron number density
relative to nucleons on the mass differences at $\sin^22\theta=1$
and for three different initial lepton asymmetries $L=10^{-10}$ (the
dashed curve) and $L=10^{-6}$ (solid curve) and $L=10^{-5}$ (the
dotted curve). \label{fig5}}
\end{minipage}
\end{center}
\end{figure}

At maximal mixing, however, bigger $L$ leads to a suppression of the
production of helium for all mass differences, and $L=10^{-5}$ is
enough to eliminate oscillations effect, i.e. to eliminate also the
constraints on oscillation parameters in the discussed BBN model
(see Fig.5).

\begin{figure}[htb]
\begin{center}
\begin{minipage}[t]{8 cm}
\epsfig{file=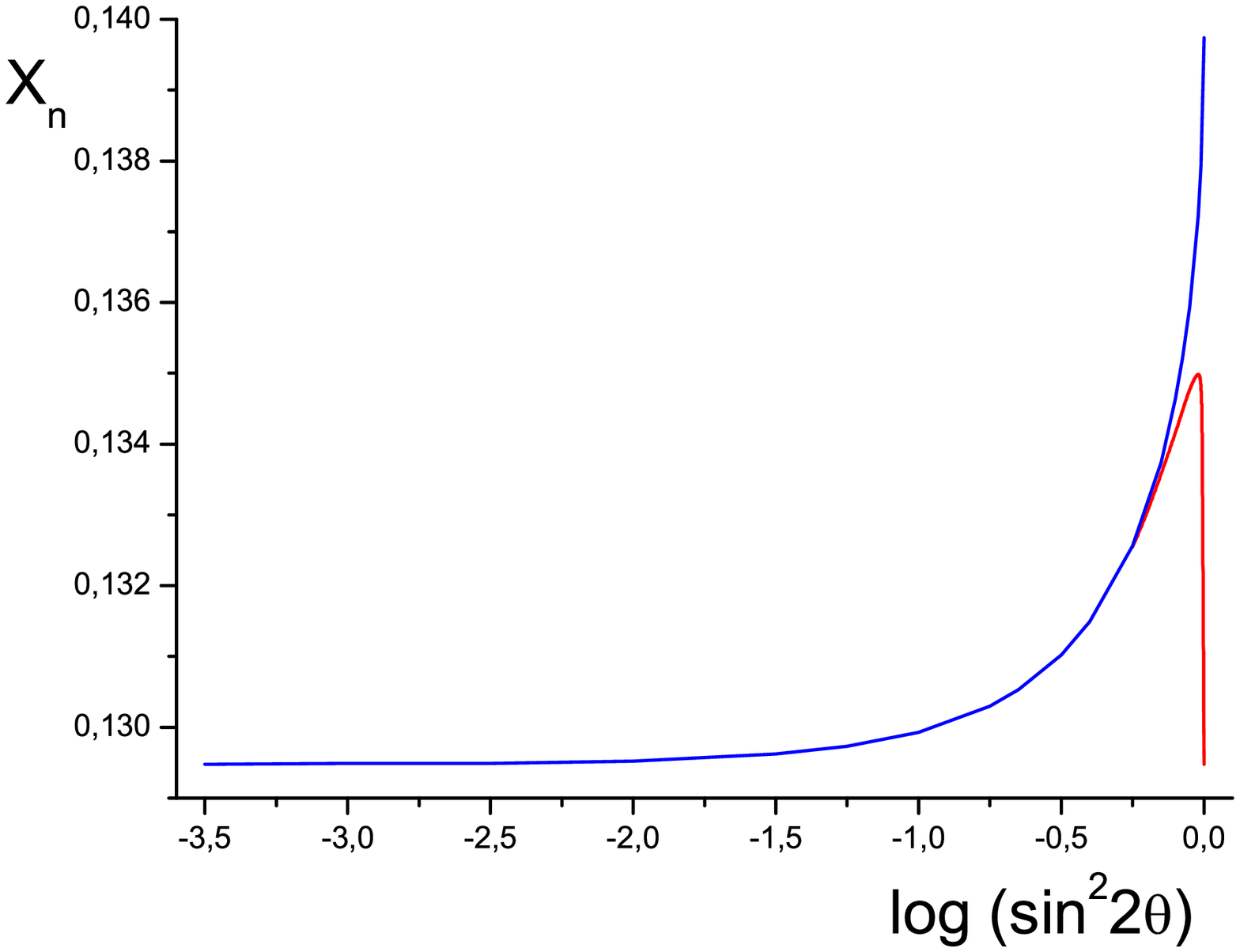,scale=0.5}
\end{minipage}
\begin{minipage}[t]{16.5 cm}
\caption{The dependence of the frozen neutron number density
relative to nucleons on the
 mixing angle at $\delta m^2=10^{-8}$ eV$^2$  and for two different
initial lepton asymmetries $L=10^{-10}$ (the dashed curve) and
$L=10^{-6}$ (solid curve). \label{fig6}}
\end{minipage}
\end{center}
\end{figure}

Finally in Fig.6 we present the dependence of the helium production
on the mixing angle at different initial $L$. Bigger $L$ leads to
decreasing the production of $^4He$ with increasing the mixing.
I.e. for fixed mass differences, $L$ relaxes the BBN constraints at
large mixings. Analysis at bigger mass differences $\delta
m^2>10^{-8}$ eV$^2$ has shown that at fixed $\delta m^2$ $L$
strengthens the constraints at small mixing. The results are in
agreement with the conclusions of ref.\cite{NPB98}, where

the  change of  BBN bounds on neutrino oscillations in the presence
of relic $L$ was studied, namely: $L$ relaxes BBN bounds at large
mixings and strengthens them at small mixings.

In conclusion, both small  asymmetry generated by neutrino
oscillations and small  relic asymmetry influences the model of  BBN
with oscillations, because the produced primordially $^4He$ in this
model feels extremely small $L$, namely $10^{-8} \ge L << 0.01$.
Hence, BBN with oscillations presents a precise leptometer.

\section{Summary}

The lepton asymmetry of the Universe may be  much bigger than the
baryon asymmetry, and hidden in the neutrino sector. Since relic
neutrino background is not yet detected, the lepton asymmetry in the
neutrino sector  may be measured/constrained just indirectly, namely
by its influence on Universe expansion, Big Bang Nucleosynthesis,
Cosmic Microwave Background, LSS, etc.

We discuss the case of small lepton asymmetry influence on the
neutrino involved processes in pre-BBN epoch, and particularly on
neutrino oscillations and BBN.


We have provided a detail numerical analysis of the interplay
between small
  lepton asymmetry
       $L << 0.01$, either relic (initially present) or dynamical
       (generated by MSW active-sterile neutrino oscillations) and
        oscillating neutrino for the case of active-sterile neutrino oscillations
       occurring after electron neutrino decoupling. The evolution of
       asymmetry growth in case of small mass differences and relatively
       big mixing angles was studied in more detail. Higher resolution
for the description of the neutrino momenta  distribution is
required for the investigation of  the asymmetry behavior  in this
oscillation parameter region. The instability region in the
oscillation parameter space, where considerable growth of $L$ takes
place, was determined numerically.
 In the case of relic lepton asymmetry we have determined  the parameter range
 for which $L$  is able to
enhance, suppress or inhibit neutrino oscillations.

Cosmological influence of such small lepton asymmetries, which do not
 have direct effect on nucleons kinetics during BBN and are invisible by
 CMB, is discussed and
 shown to be considerable.  Lepton asymmetries as small as $10^{-7}$ may
 be felt by BBN in case of neutrino oscillations.
The effect of the dynamically generated and initially present $L$ on
BBN with oscillations was studied.  Relic $L$ present during BBN,
depending on its value, may  increase, decrease  overproduction of
$Y_p$ or reduce it to the standard BBN value. Correspondingly, it
can strengthen, relax or wave out BBN constraints on oscillations.
It relaxes BBN bounds at large mixing and strengthens them at small
mixings. Large enough  $L$ alleviates BBN constraints on oscillation
parameters. In that case, instead, $L$ constraint on oscillation
parameters are derived.

Oscillations generated asymmetry at small mixing angles decreases
the production of $Y_p$ and  relaxes BBN constraints at these
angles.

The discussed model of BBN with late neutrino oscillations is
sensitive to extremely small $L$.

{\bf Acknowledgements.} I would like to thank M. Chizhov for the
overall help during the preparation of this paper.

I acknowledge the travel support  by the Bulgarian foundation
"Theoretical and Computational Physics and Astrophysics" and the EPS
grant supporting my stay at Centro Ettore Majorana during  September
16-24, 2010 where this work was finalized.


\begin{thebibliography}{99}
\itemsep -2pt

\bibitem{FVPRD96}

R. Foot, M. Thomson, R. Volkas, \Journal{\PRD}{53}{R5349}{1996}.

\bibitem{KC} D. Kirilova, M. Chizhov, {\it Neutrino96}, 478 (1996)
 D. Kirilova, M. Chizhov, \Journal{\PLB}{393}{375}{1997}.

\bibitem{friend} J. Hamann, S. Hannestad, G. Raffelt, I. Tamborra,Y.
Wang \Journal{\PRL}{105}{181301}{2010}


\bibitem{IT10} Y. Izotov T. Thuan, \Journal{\em Astrophys. J.}
{710}{L67}{2010}
\bibitem{WMAP7} E. Komatsu et al. (WMAP),\Journal{\em Astrophys. J. Suppl.}{180}{330}{2009}

\bibitem{Dolgov02} A. Dolgov, \Journal{Phys.
Rept.}{370}{333}{2002}

\bibitem{cirelli} Yi-Zen Chu, Marco Cirelli, \Journal{\PRD}{74}{085015}{2006}
A. Strumia, F. Vissani, hep-ph/0606054

\bibitem{Hannestad10} S. Hannestad \Journal{\em Prog. Part. Nucl. Phys.}{65}{185}{2010}
S. Hannestad \Journal{\em Ann. Rev. Nucl. Psrt. Sci.}{56}{137}{2006}

\bibitem{Lesgourgues&Pastor} J. Lesgourgues, S. Pastor   \Journal{\PRD}{60}
{103521}{1999};
\Journal{\em Phys. Rep.}{429}{307}{2006}

\bibitem{Shiraishi09} M. Shiraishi, K. Ichikawa, K. Ichiki, N. Sugiyama, M. Yamaguchi,
\Journal{\em JCAP}{0907}{005}{2009}

\bibitem{Simha&Steigman} Simha, G. Steigman, \Journal{\em JCAP}{0808}{011}{2008}

\bibitem{Bari} P. di Bari astro-ph/0302433 v.3 2003 Phys.Rev. D67 (2003) 127301

\bibitem{degenerate} R. Wagoner, W. Fowler, F. Hoyle, \Journal{\em Astrophys. J.}
 {148}{3}{1967};
 M. Smith, L.Kawano, R. Malaney, \Journal{\em Astrophys. J. Suppl.}{85}{219}{1993};
 H. Reeves, \Journal{\PRD}{6}{3363}{1972};
 A. Yahil, G. Beaudet, \Journal{\em Astrophys. J.}{206}{26}{1976};
G. Beaudet, P. Goret, \Journal{\em  Astron. Astrophys.}{49}{415}
{1976};
 K. Olive, D. Schramm, D. Thomas, T. Walker,
\Journal{\PRL}{B265}{239}{1991};
 H. Kang, G.
Steigman,\Journal{\NPB}{372}{494}{1992}; T. Kajino, M. Orito,
\Journal{\NPA}{629}{538C}{1998}

\bibitem{PastorPinto&Raffelt} S. Pastor, T. Pinto, G. Raffelt, \Journal{\PRL}
{102}{241302}{2009}

 \bibitem{Popa&Vasile08} L. Popa, A. Vasile \Journal{\em
 Rom. Rep. Phys.}{61}{531}{2009}

\bibitem{Schwarts&Stuke09}D. Schwarz, M. Stuke, \Journal{\em JCAP} {0911}{025}{2009}

\bibitem{NPB98}
D. Kirilova, M. Chizhov, \Journal{\em Nucl. Phys.
B}{534}{447}{1998}.


\bibitem{shi} X. Shi, \Journal{\PRD}{54}{2753}{1996}

\bibitem{NPB00} D. Kirilova, M. Chizhov, \Journal{\em Nucl. Phys.B}{591}{457}{2000}

\bibitem{Verbier} D. Kirilova, M. Chizhov, {\em in Verbier 2000,
Cosmology and particle physics}, 433 (2001), astro-ph/0101083

\bibitem{FV} R. Foot, R. R. Volkas \Journal{\PRL}{75}{4350}{1995};
\Journal{\PRD}{55}{5147}{1997}

\bibitem{FV97} R. Foot, R. R. Volkas, \Journal{\PRD}{56}{6653}{1997};\Journal{\PRD}{59}{029901}{1999}

\bibitem{dolgov&petcov} A. Dolgov, S. Hansen, S. Pastor, S.Petcov, G.Raffelt,D.Semikoz,
\Journal{\NPB}{632}{363}{2002}

 \bibitem{Serpico&Raffelt} P. Serpico, G. Raffelt, \Journal{\PRD}{71}{127301}{2005}

\bibitem{SerpicoPinto&Raffelt}  S. Pastor, T. Pinto, G. Raffelt,
 \Journal{\PRL}{102}{241302}{2009}.

\bibitem{Christel} L. Krauss, C. Lunardini, C. Smith, arXiv:1009.4666 v2

\bibitem{D04} D. Kirilova, \Journal{\em Int. J. Mod. Phys. D}{13}{831}{2004}
\bibitem{D07} D. Kirilova, \Journal{\em Int. J. Mod. Phys.
D}{16}{1197}{2007}
\bibitem{DP06} D. Kirilova, M. Panayotova, \Journal{\em JCAP}{12}{014}{2006}

\bibitem{dolgL} A. Dolgov, S. Hansen, S. Pastor,
D.Semikoz, \Journal{\em Astropart.Phys.}{14}{79}{2000}

\bibitem{Bari&Foot00} P. Di Bari, R. Foot, \Journal{\PRD}{61}{105012}{2000}

\end{thebibliography}
\end{document}